\pdfoutput=1
\documentclass[prd,nofootinbib,preprintnumbers,twocolumn,
showpacs,floatfix,superscriptaddress]{revtex4-1}

 \usepackage{auncial}
\usepackage{scrextend}
\usepackage{relsize}
\usepackage{amsmath}
\usepackage{amssymb}
\usepackage{epsfig}
\usepackage{graphicx}
\usepackage{hyperref}
\usepackage{dcolumn}   % needed for some tables
\usepackage{tabu}
\usepackage{boldline}
\usepackage{slashed}
\usepackage{multirow}
\usepackage{color}
\usepackage[normal]{subfigure}
\usepackage{rotating}
\usepackage[margin=0.9in,a4paper]{geometry}
\usepackage[table]{xcolor}
\usepackage{enumitem}
\usepackage[utf8]{inputenc}
\usepackage{colortbl}
\usepackage{array,multirow}
\definecolor{nicered}{rgb}{0.7,0.1,0.1}
\definecolor{nicegreen}{rgb}{0.1,0.5,0.1}
\definecolor{red}{rgb}{1.0, 0, 0}

\def\vev#1{\left\langle #1\right\rangle}

\def\Tr{\mbox{Tr}\,}

\def\diag{\mbox{diag}\,}

\def\gsim{\raise0.3ex\hbox{$\;>$\kern-0.75em\raise-1.1ex\hbox{$\sim\;$}}}
\def\lsim{\raise0.3ex\hbox{$\;<$\kern-0.75em\raise-1.1ex\hbox{$\sim\;$}}}
\def\mb[#1]{\mathbf{#1}}
\renewcommand{\bar}{\overline}

\definecolor{LightCyan}{rgb}{0.88,1,1}
\definecolor{piggypink}{rgb}{0.99, 0.87, 0.9}
\definecolor{applegreen}{rgb}{0.55, 0.71, 0.0}
\definecolor{darkpastelgreen}{rgb}{0.01, 0.75, 0.24}
\definecolor{green-yellow}{rgb}{0.68, 1.0, 0.18}

\newcommand{\beq}{\begin{equation}}
\newcommand{\eeq}{\end{equation}}
\newcommand{\beqa}{\begin{eqnarray}}
\newcommand{\eeqa}{\end{eqnarray}}
\newcommand{\AddrMainz}{%
PRISMA Cluster of Excellence \& Mainz Institute for Theoretical Physics, Johannes Gutenberg-Universit\"at Mainz, 55099 Mainz, Germany
}

\begin{document}
% ----------------- preprint numbers ------------------
% \begin{frontmatter}

% ------------- Title and authors ---------------------

\title{Third Family Quark-Lepton Unification at the TeV Scale}

\author{Admir Greljo}
\email{admgrelj@uni-mainz.de}
\affiliation{\normalsize\it \AddrMainz} 
\affiliation{\normalsize\it Faculty of Science, University of Sarajevo, Zmaja od Bosne 33-35, 71000 Sarajevo, Bosnia and Herzegovina}
\author{Ben A. Stefanek}
\email{bstefan@uni-mainz.de}
\affiliation{\normalsize\it \AddrMainz}

% ------------------------------------------------------
\begin{abstract}
  \noindent

We construct a model of quark-lepton unification at the TeV scale based on an $SU(4)$ gauge symmetry, while still having acceptable neutrino masses and enough suppression in flavor changing neutral currents. An approximate $U(2)$ flavor symmetry is an artifact of family-dependent gauge charges leading to a natural realization of the CKM mixing matrix.  The model predicts sizeable violation of PMNS unitarity as well as a gauge vector leptoquark $U_1^\mu = ({\bf 3}, {\bf 1}, 2/3)$ which can be produced at the LHC -- both effects within the reach of future measurements. In addition, recently reported experimental anomalies in semi-leptonic $B$-meson decays, both in charged $b \to c \tau \nu$ and neutral $b \to s \mu \mu$ currents, can be accommodated.

\end{abstract}

\maketitle

\section{Introduction}

Quark-lepton unification -- as originally suggested by Jogesh Pati and Abdus Salam~\cite{Pati:1974yy} -- is an attractive paradigm of physics beyond the Standard Model (SM). Namely, a fundamental representation of an $SU(4)$ gauge symmetry embeds a color triplet quark and a color singlet lepton (${\bf 4} = {\bf 3} \oplus {\bf 1}$). Such a construction predicts existence of an exotic particle, a gauge vector leptoquark (LQ) $U_1^\mu = ({\bf 3}, {\bf 1}, 2/3)$, which can turn a quark into a lepton and vice versa.

In this article, we entertain the possibility of quark-lepton unification at the TeV scale, motivated by the scope of present particle laboratories. The two main challenges to this idea are (i) the observed neutrino masses and (ii) the stringent constraints from flavor changing neutral currents (FCNC) in meson decays. In particular, the neutrino masses are expected to be similar in size to the masses of the up-type quarks, since the two fields are embedded in the same ${\bf 4}$ of $SU(4)$. The correct structure for a solution comes naturally in high-scale Pati-Salam models, possibly in the context of $SO(10)$ grand unification (GUT)~\cite{Babu:1992ia}, where the Majorana mass is around the GUT scale, while the Dirac mass is at the electroweak scale, leading to a seesaw mechanism~\cite{Minkowski:1977sc,Mohapatra:1979ia,Yanagida:1979as,GellMann:1980vs}. On the contrary, quark-lepton unification at scales much lower than the GUT scale (but still far beyond LHC reach) was achieved in Ref.~\cite{Perez:2013osa} using the \emph{inverse seesaw mechanism} (ISS)~\cite{Mohapatra:1986aw,Mohapatra:1986bd,GonzalezGarcia:1988rw} to generate small neutrino masses. 

Also problematic for Pati-Salam quark-lepton unification at the TeV scale are the stringent bounds on FCNC in semi-leptonic meson decays (e.g. $K_L \to \mu e$) due to gauge vector LQ exchange, pushing the LQ mass to the PeV ballpark~\cite{Valencia:1994cj,Smirnov:2007hv,Kuznetsov:2012ai,Giudice:2014tma,Smirnov:2018ske}. On the other hand, as shown recently in Ref.~\cite{DiLuzio:2017vat}, the FCNC induced by a TeV scale vector LQ can be avoided in the context of \textit{partial unification} models~\cite{Georgi:2016xhm,Diaz:2017lit} in which the SM gauge group is embedded into a larger $SU(4) \times SU(3)' \times SU(2)_L \times  U(1)'$  group (``4321"), and the (would-be) SM fermions are charged only under the ``321" part. The LQ couplings to SM fermions are generated via mass mixing with extra vector-like fermions charged under $SU(4)$, where the largest LQ interactions are taken to be with the third family fermions as allowed by the low energy flavor data. Note that this construction does not have a neutrino mass problem since the (would-be) SM quarks and leptons are not unified in ${\bf 4}$ of $SU(4)$.\footnote{As shown in~\cite{DiLuzio:2017vat}, the ``$4321$" model is the first UV complete gauge model to coherently address a set of experimental anomalies recently reported in semi-leptonic $B$-meson decays~\cite{Lees:2013uzd,Hirose:2016wfn,Aaij:2015yra,Aaij:2014ora,Aaij:2017vbb,Aaij:2013qta,Aaij:2015oid}, utilizing the vector LQ representation $U_1^\mu = ({\bf 3}, {\bf 1}, 2/3)$. See also recent activities in Refs.~\cite{Buttazzo:2017ixm,Assad:2017iib,Calibbi:2017qbu,Bordone:2017bld,Barbieri:2017tuq,Blanke:2018sro}.} 

Building on this work, the authors of Ref.~\cite{Bordone:2017bld} introduce family-dependent gauge interactions -- Pati-Salam for every family ($PS^3$) -- achieving a TeV scale vector LQ dominantly coupled to the third family while still having quarks and leptons unified into a ${\bf 4}$ of $SU(4)$. Scalar link fields are introduced to break the gauge symmetry down to the SM. This is done in several steps with very hierarchical vacuum expectation values (VEVs) ranging from $1$~TeV up to (at least) $10^3$~TeV -- a construction which is presumably responsible for the peculiar quark masses and mixing in the SM. However, the aforementioned neutrino mass problem is set aside noting that, in principle, one could fine tune the contributions of the two Higgs fields, both of which are $\mathcal{O}(v_{ \rm EW})$ where $v_{\rm EW}\approx246$~GeV. 

Also relevant to this article is the idea of Ref.~\cite{Chivukula:2013kw}, where the authors consider an extended color symmetry $SU(3)_{12} \times SU(3)_3 \to SU(3)_c$~, where the first two quark families are charged under $SU(3)_{12}$, and the third family is charged under $SU(3)_3$. An approximate $U(2)$ flavor symmetry~\cite{Barbieri:2011ci} is obtained accidentally as an artifact of the gauge representation choices. The leading $U(2)$ breaking spurion is generated by integrating out a  weak doublet vector-like quark.

Building on the work of the aforementioned Refs.~\cite{Perez:2013osa,Chivukula:2013kw,DiLuzio:2017vat,Bordone:2017bld}, we construct a model of TeV scale quark-lepton unification based on the ``$4321$" gauge group with the third family charged under ``$421$", and the light families under ``$321$". As a consequence, the model possesses an accidental approximate $U(2)$ flavor symmetry which is softly broken by a weak doublet vector-like fermion representation. In addition, SM singlet fermions are introduced in order to implement the inverse seesaw mechanism and generate acceptable neutrino masses and mixings without a fine-tuning problem. When the vector-like fermion is integrated out, this model is the low-energy limit of Ref.~\cite{Bordone:2017bld}, apart from the neutral lepton sector. The model is UV complete and renormalizable, and the heaviest states are not far above the TeV scale.  Therefore, unlike high scale models of quark-lepton unification, our model does not introduce a severe problem with the stabilization of the electroweak scale.

%\admir{This model predicts the gauge boson spectrum of Ref.~\cite{DiLuzio:2017vat}, but is instead similar to the low-energy limit of Ref.~\cite{Bordone:2017bld} except for the neutral lepton sector.}

%%%%%%%%%%%%%%%%%
\section{Model Basics}
%%%%%%%%%%%%%%%%%

\subsection{Gauge Symmetry and Breaking Structure} 
We consider here the ``$4321$" gauge group $G \equiv SU(4) \times SU(3)' \times SU(2)_L \times U(1)'$.
We label the respective gauge fields as $H^\alpha_\mu, G'^a_\mu, W^i_\mu, B'_\mu$, the gauge couplings as 
$g_4, g_3, g_2, g_1$, and the generators as $T^\alpha, T^a, T^i, Y'$ with indices $\alpha = 1, \dots, 15$, $a = 1, \dots, 8$, and $i=1,2,3$.  
The generators are normalized such that $\Tr [T^{A}T^{B}]= \frac{1}{2} \delta^{AB}$ in the fundamental representation. The ``4321" gauge group $G$ contains the SM gauge group $G_{\rm SM} = SU(3)_c \times SU(2)_L \times U(1)_Y$ as a subgroup. Specifically, color is embedded as
$SU(3)_c = \left[ SU(3)_4 \times SU(3)' \right]_{\rm diag}$ and hypercharge is embedded as $U(1)_Y = \left[ U(1)_4 \times U(1)' \right]_{\rm diag}$, 
where $SU(3)_4 \times U(1)_4 \subset SU(4)$. 
\begin{table}[ht!]
\renewcommand{\arraystretch}{1.1}
\begin{center}
\begin{tabular}{|c|c|c|c|c||c|c|}
\hline 
  \multicolumn{7}{|c|}{ \rm Scalar Fields} \\ \hline
  \multicolumn{5}{|c||}{Gauge} & \multicolumn{2}{|c|}{Global}\\ \hline
 {\rm Field } & $SU(4)$    &   $SU(3)'$   & $SU(2)_{L}$ & $U(1)'$ & $U(1)_{B'}$ & $U(1)_{L'}$\\ \hline
  $H$  & {\bf 1}  &  {\bf 1}  &  {\bf 2}  &  1/2 & 0 & 0      \\ 
  $\Phi$  & {\bf 15}  &  {\bf 1}  &  {\bf 2}  &  1/2 & 0 & 0     \\ 
  $\Omega_{3}$  & $\overline{{\bf 4}}$  &  {\bf 3}  &  {\bf 1}  &  1/6  & 1/12 & -1/4     \\ 
  $\Omega_{1}$  & $\overline{{\bf 4}}$  &  {\bf 1}  &  {\bf 1}  &  -1/2  & -1/4 & 3/4     \\ 
  \hline
\end{tabular}
\end{center}
\caption{Scalar sector of the model.} 
\label{tab:scalar_tab}
\end{table}
Spontaneous symmetry breaking of $G \to G_\text{SM}$ occurs when the scalars  $\Omega_3$ and $\Omega_1$ shown in Table~\ref{tab:scalar_tab} acquire vacuum expectation values. The proper $G \to G_\text{SM}$ breaking is achieved by following VEV configurations~\cite{Diaz:2017lit,DiLuzio:2017vat}
\beq  
\label{vevconf}
\vev{\Omega_3} = 
\left(
\begin{array}{ccc}
\tfrac{v_3}{\sqrt{2}} & 0 & 0 \\
0 & \tfrac{v_3}{\sqrt{2}} & 0 \\ 
0 & 0 & \tfrac{v_3}{\sqrt{2}} \\
0 & 0 & 0
\end{array}
\right) \, , \ \ 
\vev{\Omega_1} = 
\left(
\begin{array}{c}
0 \\ 
0 \\ 
0 \\
\tfrac{v_1}{\sqrt{2}}
\end{array}
\right) \, .
\eeq
In the $G$-symmetry broken phase, we have $Y = \sqrt{\tfrac{2}{3}}T^{15} + Y'$, where $T^{15} = \tfrac{1}{2 \sqrt{6}} \text{diag}(1, 1, 1, -3)$. The $G$-symmetry breaking scalar representations decompose under $G_\text{SM}$ as 
$\Omega_3 = ({\bf 8},{\bf 1},0) \oplus ({\bf 1},{\bf 1},0) \oplus ({\bf 3},{\bf 1},2/3)$ and 
$\Omega_1 = ({\bf \bar 3},1,-2/3) \oplus ({\bf 1},{\bf 1},0)$. 
In the unitary gauge, the physical scalar degrees of freedom are: a real color octet, three real singlets, and a complex triplet leptoquark.

Additionally, there are three massive gauge bosons $U_{1}$, $g'$, and $Z'$ which belong to the coset group
$G / G_{\rm SM}$. They transform as $U_{1} = ({\bf 3},{\bf 1},2/3)$, $g' = ({\bf 8},{\bf 1},0)$ and $Z' = ({\bf 1},{\bf 1},0)$ under $G_{\rm SM}$ and have masses~\cite{Diaz:2017lit,DiLuzio:2017vat}
\begin{align}
\label{MV}
m_{U_{1}} &= \tfrac{1}{2} g_4 \sqrt{v_1^2 + v_3^2} \, , \\
\label{Mgp}
m_{g'} &= \tfrac{1}{\sqrt{2}} v_3 \sqrt{g_4^2 + g_3^2} 
\, , \\
\label{MZp}
m_{Z'} &= \tfrac{1}{2} \sqrt{\tfrac{3}{2}} \sqrt{g_4^2 + \tfrac{2}{3} g_1^2} \sqrt{v_1^2 + \tfrac{1}{3} v_3^2} 
\, .
\end{align}
Expressions for $U_{1}$, $g'$, $Z'$, and the SM gauge bosons and gauge couplings in terms of the original gauge fields and gauge couplings of $G$ can be found in Ref.~\cite{DiLuzio:2017vat}. A benchmark point matching the SM gauge couplings to the gauge couplings of $G$ at $\mu = 2$ TeV yields $g_{4} = 3$, $g_{3} = 1.08$, and $g_{1} = 0.365$.

The final breaking is electroweak symmetry breaking, $G_\text{SM} \rightarrow SU(3)_c \times U(1)_{\rm EM}$, obtained when the Higgs doublet $H = ({\bf 1},{\bf 1},{\bf 2},1/2)$ of $G$ acquires a VEV $\vev{H_0} = v_{H}/\sqrt{2}$. Additionally, there is an $SU(4)$ adjoint scalar $\Phi \equiv \Phi^\alpha T^\alpha= ({\bf 15},{\bf 1},{\bf 2},1/2)$ which contains weak doublets of several kinds: a color octet, two color triplets and a color singlet $\Phi^{15}$ (another Higgs doublet). In what follows, we assume that only $\vev{\Phi^{15}_0} = v_{\Phi}/\sqrt{2}$ develops a VEV and contributes to electroweak symmetry breaking. 
The scalar potential of the model can naturally generate the aforementioned VEVs and symmetry breaking pattern~\cite{inpreparation}.
We note that apart from the addition of the $SU(4)$ adjoint scalar $\Phi$, the bosonic sector of the model is identical to that of Ref.~\cite{DiLuzio:2017vat}.

%%%%%%%%%%%%%%%%%
\subsection{Matter Content} 
\label{mat_cont}
%%%%%%%%%%%%%%%%%
The would-be light family SM fermion fields (when neglecting the mixing discussed below), are charged under the $SU(3)' \times SU(2)_L \times U(1)'$ subgroup, but are singlets of $SU(4)$. Let us denote them as: $q'^i_L = ({\bf 1},{\bf 3},{\bf 2},1/6)$, $u'^i_R = ({\bf 1},{\bf 3},{\bf 1},2/3)$, 
$d'^i_R = ({\bf 1},{\bf 3},{\bf 1},-1/3)$,  $\ell'^i_L = ({\bf 1},{\bf 1},{\bf 2},-1/2)$, and $e'^i_R = ({\bf 1},{\bf 1},{\bf 1}, -1)$. We label these representations as dominantly light family SM fermions and note that they come in two copies of flavor ($i = 1,2$). Being $SU(4)$ singlets, they do not couple with the vector leptoquark $U_{1}^\mu$ directly.

In contrast, the would-be third family SM fermion fields are charged as fundamentals under $SU(4)$, in addition to carrying charge under $SU(2)_L \times U(1)'$. We denote them as: $\psi_{L}= ({\bf 4},{\bf 1},{\bf 2},0)$, $\psi_{R}^{u}= ({\bf 4},{\bf 1},{\bf 1},1/2)$, and $\psi_{R}^{d}=({\bf 4},{\bf 1},{\bf 1},-1/2)$. The dominantly third family SM fermions are embedded into these representations as $(\psi_{L})^{T} = ( q_{L}'^{3} \, \, \,  \ell_{L}'^{3}) $,  $(\psi_{R}^{u})^{T} = ( u_{R}'^{3} \, \, \,  \nu_{R}'^{3})$, and  $(\psi_{R}^{d})^{T} = ( d_{R}'^{3} \, \, \,  e_{R}'^{3})$. This field content is summarized in Table \ref{tab:fieldcontent}. Unlike the light family fermions, the dominantly third family SM fermions couple directly to the vector leptoquark $U_{1}^\mu$ via gauge interactions. 

\begin{table}[ht]
\renewcommand{\arraystretch}{1.1}
\begin{center}
\begin{tabular}{|c|c|c|c|c||c|c|}
\hline 
\multicolumn{7}{|c|}{ \rm Dominantly Light Family SM Fermions} \\ \hline
 \multicolumn{5}{|c||}{Gauge} & \multicolumn{2}{|c|}{Global}\\ \hline
 {\rm Field } & $SU(4)$    &   $SU(3)'$   & $SU(2)_{L}$ & $U(1)'$ & $U(1)_{B'}$ & $U(1)_{L'}$\\ \hline
 $q_{L}'^{i}$  & {\bf 1}  &  {\bf 3}  &  {\bf 2}  & 1/6  & 1/3 & 0     \\ 
 $u_{R}'^{i}$  & {\bf 1}  &  {\bf 3}  &  {\bf 1}  & 2/3  & 1/3 & 0     \\ 
 $d_{R}'^{i}$  & {\bf 1}  &  {\bf 3}  &  {\bf 1}  & -1/3 & 1/3 & 0      \\ 
 $\ell_{L}'^{i}$  & {\bf 1}  &  {\bf 1}  &  {\bf 2}  & -1/2  & 0 & 1     \\ 
 $e_{R}'^{i}$  & {\bf 1}  &  {\bf 1}  &  {\bf 1}  & -1  & 0 & 1     \\ 
  \hline \multicolumn{7}{c}{ } \\ \hline
 \multicolumn{7}{|c|}{ \rm Dominantly Third Family SM Fermions} \\ \hline
 \multicolumn{5}{|c||}{Gauge} & \multicolumn{2}{|c|}{Global}\\ \hline
 {\rm Field } & $SU(4)$    &   $SU(3)'$   & $SU(2)_{L}$ & $U(1)'$ & $U(1)_{B'}$ & $U(1)_{L'}$\\ \hline
$\psi_{L}$  & {\bf 4}  & {\bf 1}  &  {\bf 2}  & 0  & 1/4 & 1/4     \\ 
$\psi^{u}_{R}$  & {\bf 4}  &  {\bf 1}  &  {\bf 1}  & 1/2 & 1/4 & 1/4     \\ 
$\psi^{d}_{R}$  & {\bf 4}  &  {\bf 1}  &  {\bf 1}  & -1/2 &1/4 &1/4     \\ 
  \hline
\end{tabular}
\end{center}
\caption{Dominantly SM fermion content of the model in the $G$-symmetric phase. The flavor index $i = 1,2$ runs over the 1st and 2nd family fermions, while the third family is embedded in $\psi_{L}$, $\psi^{u}_{R}$, and $\psi^{d}_{R}$.
\label{tab:fieldcontent}}
\end{table}

In order to generate mixing between the third and light family fermions, we introduce a vector-like fermion representation $\chi_{L,R} = ({\bf 4},{\bf 1},{\bf 2},0)$, shown in Table \ref{tab:VLF_tab}. This representation decomposes under the SM as  $\chi_{L,R}^T \equiv (Q'_{L,R}, L'_{L,R})$, where $Q'_{L,R}$ and $L'_{L,R}$ are vector-like partners of the SM quark and lepton doublets, respectively. The left-handed field $\chi_{L}$ couples to the right-handed dominantly third family SM fermions $\psi_{R}^{u}$ and $\psi_{R}^{d}$ via a Higgs insertion. The right-handed field $\chi_{R}$ couples to the left-handed dominantly light family SM quark doublets via $\Omega_{3}$ insertions and to the left-handed dominantly light family SM lepton doublets via $\Omega_{1}$ insertions.
\begin{table}[h!]
\renewcommand{\arraystretch}{1.1}
\begin{center}
\begin{tabular}{|c|c|c|c|c||c|c|}
\hline 
 \multicolumn{7}{|c|}{ \rm New Vector-like Fermions} \\ \hline
 \multicolumn{5}{|c||}{Gauge} & \multicolumn{2}{|c|}{Global}\\ \hline
 {\rm Field } & $SU(4)$    &   $SU(3)'$   & $SU(2)_{L}$ & $U(1)'$ & $U(1)_{B'}$ & $U(1)_{L'}$\\ \hline
$\chi_{L,R}$  & {\bf 4}  &  {\bf 1}  &  {\bf 2}  & 0  & 1/4 & 1/4     \\ 
  \hline
\end{tabular}
\end{center}
\caption{Vector-like fermion representation.} 
\label{tab:VLF_tab}
\end{table}

Since the dominantly third family SM quarks and leptons are unified into $SU(4)$ multiplets, we get the interesting prediction that $m_{t}' = m_{\nu_{\tau}}'$ and $m_{b}' = m_{\tau}'$ if they receive mass only from the Higgs field. While this approximation works quite well for the bottom quark and tau lepton, the prediction that the top quark and tau neutrino must have the same mass is extremely inconsistent with experimental data. If the dominantly third family SM fermions also receive contributions to their masses from the VEV of $\Phi^{15}$, then there are four independent Yukawa couplings and correct masses for all third family SM fermions can be achieved. However, a large fine-tuning is required to arrange a cancellation between the two terms contributing to the tau neutrino mass in order to obtain an experimentally acceptable value. This fine-tuning problem for neutrino masses in low scale $SU(4)$ quark-lepton unification models was solved by Ref.~\cite{Perez:2013osa} by adding singlet fermions to implement the inverse seesaw mechanism. Here, we follow this prescription and introduce two right-handed dominantly light family SM neutrinos $\nu_{R}'^{i}$ and three new right-handed fermions $S^{a}_{R}$ which are singlets under $G$. This extension of the fermion content is summarized in Table \ref{tab:singlet_tab} and we discuss the details of the ISS mechanism in Section \ref{lep_sec}.

\begin{table}[h!]
\renewcommand{\arraystretch}{1.1}
\begin{center}
\begin{tabular}{|c|c|c|c|c||c|c|}
\hline 
  \multicolumn{7}{|c|}{ \rm Right Handed Singlet Fermions} \\ \hline
  \multicolumn{5}{|c||}{Gauge} & \multicolumn{2}{|c|}{Global}\\ \hline
 {\rm Field } & $SU(4)$    &   $SU(3)'$   & $SU(2)_{L}$ & $U(1)'$ & $U(1)_{B'}$ & $U(1)_{L'}$\\ \hline
 $\nu_{R}'^{i}$  & {\bf 1}  &  {\bf 1}  &  {\bf 1}  & 0  & 0 & 1     \\ 
 $S_{R}^{a}$  & {\bf 1}  &  {\bf 1}  &  {\bf 1}  & 0 & 0 & -1      \\ 
  \hline
\end{tabular}
\end{center}
\caption{Fermion singlets. The index $i=1,2$ ($a=1,2,3$) is a flavor index for the gauge singlet fermion $\nu'_R$ ($S_R$).} 
\label{tab:singlet_tab}
\end{table}

In addition, there are accidental global symmetries $U(1)_{B'}$ and $U(1)_{L'}$,  whose action on the matter fields are displayed in the last two columns of the first four tables. 
The VEVs of $\Omega_3$ and $\Omega_1$ spontaneously break both the gauge and global symmetries, leaving two new global $U(1)$'s unbroken: 
$B = B'+\frac{1}{\sqrt{6}}T^{15}$ and $L = L'-\sqrt{\frac{3}{2}}T^{15}$.
For SM particles, these unbroken $U(1)$'s correspond to ordinary baryon and lepton number, respectively. These symmetries protect proton stability and make the active neutrinos massless. {As will be discussed later on, a soft breaking of $U(1)_{L}$ will lead to tiny neutrino masses in the context of the inverse seesaw mechanism. }

%%%%%%%%%%%%%%%%%
\section{Yukawa Interactions} 
%%%%%%%%%%%%%%%%%
Let us define $\Psi^T_L=(\psi_L, \chi_L)$ and write the Lagrangian containing Yukawa interactions and mass terms as $\mathcal{L}_{\rm Yuk} = \mathcal{L}_{12} + \mathcal{L}_{3 \chi} + \mathcal{L}_{\nu}$, where
\begin{equation} \label{YukL}
\begin{split}
\mathcal{L}_{12} =& - \overline{q}_{L}' \,Y_{u}  \,\widetilde{H} \,u_{R}' - \overline{q}_{L}' \,Y_{d} \, H \,d_{R}' \\
&-  \overline{\ell}_{L}' \,Y_{\nu}\, \widetilde{H}\,  \nu_{R}' - \overline{\ell}_{L}' \,Y_{e}\, H \,e_{R}' +{\rm h.c.}~, \\
\mathcal{L}_{3 \chi} =&  - \bar {q}'_L \lambda_q \, \Omega_3^T \chi_R - \bar {\ell}'_L \lambda_\ell \, \Omega_1^T \chi_R - \bar \Psi_L {\bf m}\,  \chi_R~\\
& - \overline{\Psi}_{L} \, {\bf y}_{H}^{u} \widetilde{H} \psi_{R}^{u} - \overline{\Psi}_{L}\, {\bf y}_{H}^{d} H \psi_{R}^{d}\\
& -  \overline{\Psi}_{L} \, {\bf y}_{\Phi}^{u} \, \widetilde{\Phi} \, \psi_{R}^{u} - \overline{\Psi}_{L}\, {\bf y}_{\Phi}^{d} \,  \Phi  \, \psi_{R}^{d} + \text{h.c.}~,\\
\mathcal{L}_{\nu} =& - \Omega_{1}^T\, \overline{S_{R}^{c}} \,  \lambda_{R}\psi_{R}^{u} - \overline{S_{R}^{c}} \, M_{R} \, \nu_{R}' ~\\
&-\frac{1}{2}  \overline{S^{c}_{R}}  \,\mu_{S} \, S_{R} - \frac{1}{2}  \overline{\nu'^{c}_{R}}\,\mu_{R}\,\nu_{R}' + {\rm h.c.}~.
\end{split}
\end{equation}
Here, we have defined $\widetilde{H} \equiv i \sigma^2 H^*$, $\widetilde{\Phi} \equiv T^\alpha (i\sigma^2 \Phi^{\alpha *})$, ${\bf m}^{T} = (m_{\rm mix},m_\chi)$, $ ({\bf y}_{H,\Phi}^{u})^{T} = (y_{H,\Phi}^u,\lambda_{H, \Phi}^{u})$ and $ ({\bf y}_{H,\Phi}^{d})^{T} = (y_{H,\Phi}^d,\lambda_{H, \Phi}^{d})$.

Without $\mathcal{L}_{\rm Yuk}$, the global flavor symmetry of the model is  $\mathcal{F} = \mathcal{F}_{3\chi} \times \mathcal{F}_{12} \times \mathcal{F}_{S}$, where $\mathcal{F}_{3\chi} = U(2)_{\Psi_L}  \times U(1)_{\psi_{R}^{u}} \times U(1)_{\psi_{R}^{d}} \times U(1)_{\chi_{R}}$ is the flavor symmetry of the dominantly third family SM fermions and the new vector-like fermion $\chi_{L,R}$, $\mathcal{F}_{12} =  U(2)_{q'} \times U(2)_{u'} \times U(2)_{d'} \times U(2)_{\ell '} \times U(2)_{e'} \times U(2)_{\nu'}$ is the flavor symmetry of the dominantly light family SM fermions, and $\mathcal{F}_{S} = U(N_{S})_{S} $ is the flavor symmetry of the right-handed singlet fields $S_{R}^a$. The breaking $\mathcal{F} \rightarrow U(1)_{B'}$ of the flavor group occurs when $\mathcal{L}_{\rm Yuk}$ is present. We can use the broken $\mathcal{F}_{12}$ flavor symmetry start in a basis in which $Y_{d}$ and $Y_{e}$ are real and diagonal, $Y_{u} = V^{\dagger} Y_{u}^{\rm diag}$, and $Y_{\nu} = U Y_{\nu}^{\rm diag}$. Here, $V$ and $U$ are orthogonal $2 \times 2$ matrices, with $V$ approximately the Cabibbo matrix. Since $\mu_{S}$ and $\mu_{R}$ softly break $U(1)_{L'}$, the group $\mathcal{F}_{S}$ is fully broken and can be used to make $\mu_{S}$ real and diagonal.  

The broken $\mathcal{F}_{3\chi}$ symmetry allows us to choose $m_{\rm mix} =0$, $m_{\chi}$ real, and fix the phases of $y_{H}^{u}$ and $y_{H}^{d}$.  The remaining broken symmetry includes the dominantly third family fermion number $U(1)_{3} \subset \mathcal{F}_{3\chi}$ and the dominantly light family fermion number $U(1)_{12} \subset \mathcal{F}_{12}$.
We will later use the $U(1)_{3}$ to adjust the phase of $\lambda_{H}^{d}$ and the $U(1)_{12}$ to choose one component of $\lambda_{q}^{T} \equiv (\lambda_{q}^{(1)} , \, \lambda_{q}^{(2)})$ to be real. If $m_{\chi}\gtrsim$ TeV, $\chi$ can be integrated out to generate dimension-5 operators which mix the third and light family fermions, closely approaching the setup of~\cite{Bordone:2017bld}
{\footnotesize
\begin{equation} \label{Ld5}
\begin{split}
\mathcal{L}_{d5} &= \frac{\lambda_{q}}{m_{\chi}} \left( \lambda_{H}^{u} \, \bar q'_{L} \Omega_{3}^T \widetilde{H} \, \psi_R^u + \lambda_{\Phi}^{u} \, \bar q'_{L} \Omega_{3}^T \widetilde{\Phi}  \, \psi_R^u \right)\\ 
&  +\frac{\lambda_{q}}{m_{\chi}} \left(  \lambda_{H}^{d} \, \bar q'_{L} \Omega_{3}^T H \, \psi_R^d +\lambda_{\Phi}^{d} \, \bar q'_{L} \Omega_{3}^T \Phi \, \psi_R^d  \right)\\
& +\frac{\lambda_{\ell}}{m_{\chi}} \left( \lambda_{H}^{u} \, \bar \ell'_{L} \Omega_{1}^T \widetilde{H} \, \psi_R^u + \lambda_{\Phi}^{u} \, \bar \ell'_{L} \Omega_{1}^T \widetilde{\Phi} \, \psi_R^u \right) \\
& +\frac{\lambda_{\ell}}{m_{\chi}}\left( \lambda_{H}^{d} \, \bar \ell'_{L} \Omega_{1}^T H \, \psi_R^d+\lambda_{\Phi}^{d} \, \bar \ell'_{L} \Omega_{1}^T  \Phi \, \psi_R^d \right) + {\rm h.c.}~.
\end{split}
\end{equation}}
After electroweak symmetry breaking, the dominantly third family SM fermions receive the following masses~\cite{Perez:2013osa}
{\footnotesize
\begin{align}
\label{mup}
m_{t}' &= \frac{v_{\rm EW}}{\sqrt{2}} \left( y_{H}^{u} \cos\beta + \frac{1}{2\sqrt{6}}y_{\Phi}^{u} \sin\beta\right) \, ,  \\
\label{mvp}
m_{\nu_{\tau}}' &= \frac{v_{\rm EW}}{\sqrt{2}} \left( y_{H}^{u} \cos\beta - \frac{3}{2\sqrt{6}}y_{\Phi}^{u} \sin\beta\right) \, , \\
\label{mdp}
m_{b}' &= \frac{v_{\rm EW}}{\sqrt{2}} \left( y_{H}^{d} \cos\beta + \frac{1}{2\sqrt{6}}y_{\Phi}^{d} \sin\beta\right) \, , \\
\label{mep}
m_{\tau}' &= \frac{v_{\rm EW}}{\sqrt{2}} \left( y_{H}^{d} \cos\beta - \frac{3}{2\sqrt{6}}y_{\Phi}^{d} \sin\beta\right) \, ,
\end{align}}
\!\!where we have defined $v_{\rm EW}^{2} = v_{H}^{2} + v_{\Phi}^{2}$ and $\tan\beta =  v_{\Phi}/v_{H}$. Since we have the freedom to fix the phases of $y_{H}^{u}$ and $y_{H}^{d}$, we will choose them such that the linear combinations which comprise $m_{t}'$ and $m_{b}'$ are real. We also simplify the light and third family fermion mixing coefficients from $\mathcal{L}_{d5}$ by defining
{\footnotesize
\begin{align}
\label{fup}
f_{u} &= \frac{v_{3}v_{\rm EW}}{2m_{\chi}} \left( \lambda_{H}^{u} \cos\beta + \frac{1}{2\sqrt{6}}\lambda_{\Phi}^{u} \sin\beta \right)\, ,  \\
\label{fvp}
f_{\nu} &= \frac{v_{1}v_{\rm EW}}{2m_{\chi}} \left( \lambda_{H}^{u} \cos\beta - \frac{3}{2\sqrt{6}}\lambda_{\Phi}^{u} \sin\beta \right) \, , \\
\label{fdp}
f_{d} &= \frac{v_{3}v_{\rm EW}}{2m_{\chi}} \left( \lambda_{H}^{d} \cos\beta + \frac{1}{2\sqrt{6}}\lambda_{\Phi}^{d} \sin\beta \right) \, , \\
\label{fep}
f_{e} &= \frac{v_{1}v_{\rm EW}}{2m_{\chi}}  \left( \lambda_{H}^{d} \cos\beta - \frac{3}{2\sqrt{6}}\lambda_{\Phi}^{d} \sin\beta \right)\, .
\end{align}}

\subsection{Quark Sector}

The quark mass matrices have the same structure as in Ref.~\cite{Chivukula:2013kw}
\begin{equation}
\mathcal{M}_{d} = 
\renewcommand{\arraystretch}{1.3}
\left(
\begin{matrix}
\frac{v_{H}}{\sqrt{2}} Y_{d}^{\rm diag}  & -f_{d}\, \lambda_{q}  \\
0  &  m'_{b} \\
\end{matrix}
\right) \,,
\end{equation}

\begin{equation}
\mathcal{M}_{u} = 
\renewcommand{\arraystretch}{1.3}
\left(
\begin{matrix}
\frac{v_{H}}{\sqrt{2}} V^{\dagger}Y_{u}^{\rm diag}  &    -f_{u}\, \lambda_{q}   \\
0  &  m'_{t}  \\
\end{matrix}
\right) \, .
\end{equation}

These $3\times3$ matrices can be diagonalized by bi-unitary rotations of the form $ \mathcal{M}_{f}^{\rm diag} = V_{L}^{f}\mathcal{M}_{f} V_{R}^{f \dagger}$ with $f = u,d$. With this convention, the Cabibbo-Kobayashi-Maskawa (CKM) matrix is defined as $V_{\rm CKM} = V_{L}^{u}V_{L}^{d\dagger}$. Diagonalizing $\mathcal{M}_{u}$ and $\mathcal{M}_{d}$ assuming $v_{H} Y_{u,d}^{\rm diag}  \ll m_{\chi}|\lambda_{q}|, \, f_{u,d}, \, m_{t,b}'  \ll m_{\chi}$, the CKM matrix to leading order is
{\footnotesize
\begin{equation}
V_{\rm CKM} = 
\renewcommand{\arraystretch}{1.3}
\left(
\begin{matrix}
V_{ud}  & V_{us} & F \left(V_{ud} \lambda_{q}^{(1)} + V_{us} \lambda_{q}^{(2)}\right)  \\
V_{cd} & V_{cs}  &  F \left(V_{cd} \lambda_{q}^{(1)} + V_{cs} \lambda_{q}^{(2)}\right)   \\
-\left(F \, \lambda_{q}^{(1)}\right)^{*} & -\left(F \,  \lambda_{q}^{(2)}\right)^{*}  &  1  \\
\end{matrix}
\right) \, ,
\end{equation}}
\!\!where we have defined
\begin{equation}
F = \frac{f_{u}}{m_{t}'}- \frac{f_{d}}{m_{b}'} \, .
\end{equation}
As mentioned previously, we have the freedom remaining to choose $\lambda_{q}^{(2)}$ real and to fix the phase of $\lambda_{H}^{d}$ such that $  {\rm Im} (f_{u})/ m_{t}' = {\rm Im} (f_{d})/m_{b}' $, making $F$ real.  Comparing to the Wolfenstein parameterization in Ref.~\cite{Patrignani:2016xqp}, the CKM matrix can be fit by: $V_{td} =A\lambda^{3}(1-\rho-i\eta) = -F(\lambda_{q}^{(1)})^{*} =0.0080 - i\, 0.0033$ and $V_{ts} = -A\lambda^{2} = -F\lambda_{q}^{(2)} = -0.041$. 

The $U(2)_{q'} \times U(2)_{u'} \times U(2)_{d'}$ flavour symmetry of the quark sector is softly broken by the spurion bi-doublets $Y_u  \sim ({\bf 2},{\bf \bar 2},{\bf 1})$,  $Y_d \sim ({\bf 2},{\bf 1},{\bf \bar 2})$, and a single spurion doublet ${\bf V}^{(i)}=\frac{v_3}{m_\chi} \lambda_{q}^{(i)} \sim ({\bf 2},{\bf 1},{\bf 1})$ which is entirely responsible for the communication of the third to light generations. This setup nicely reproduces the Minimal $U(2)$ picture of quark masses and mixings proposed in Ref.~\cite{Barbieri:2011ci}.
The smallness of the leading breaking spurion doublet ${\bf V}^{(i)}$ can be understood as a consequence of large $m_\chi$ or perhaps small $\lambda_{q}^{(i)}$, which is the only coupling violating the light family quark number.

\subsection{Lepton Sector}
\label{lep_sec}
As mentioned in Section~\ref{mat_cont}, we introduced two right-handed dominantly light family SM neutrinos $\nu_{R}'^{i}$ and three new right-handed fermions $S^{a}_{R}$ which are singlets under $G$ in order to avoid fine-tuning in  Eq.~\eqref{mvp}. {To see how this is achieved, we define $n_{L}^{T} = (\nu_{L}' \,\, \nu_{R}'^{c} \,\, S_{R}^{c})$, where $\nu'$ contains the light and third family neutrinos and $S$ contains all its flavors.}  When all scalars receive VEVs, the neutrino mass Lagrangian can be written as
\begin{equation}
\mathcal{L}_{\nu} = -\frac{1}{2} \overline{n}_{L} \mathcal{M}_{\nu} n_{L}^{c} + \textrm{h.c.}\, .
\end{equation}
The neutrino mass matrix $\mathcal{M_{\nu}}$ is a $9 \times 9$ matrix of the form
\begin{equation}
\mathcal{M}_{\nu} = 
\renewcommand{\arraystretch}{1.3}
\left(
\begin{matrix}
0 &   M_{\nu}^{D} & 0  \\
(M_{\nu}^{D})^{T}  &  \widetilde{\mu}_{R} & \widetilde{M}_{R}^{T}  \\
0 & \widetilde{M}_{R} & \mu_{S} \\
\end{matrix}
\right) \, ,
\label{eq:ISS_matrix}
\end{equation}
where we have defined $ \widetilde{M}_{R} \equiv (M_{R} \,\,\, \frac{v_{1}}{\sqrt{2}}\lambda_{R})$. The $3\times 3$ matrix $\widetilde{\mu}_{R}$ contains $\mu_{R}$ as the upper left $2\times 2$ block and has zeros elsewhere. The Dirac mass matrix $M_{\nu}^{D}$ is a $3 \times 3$ matrix of the form
\begin{equation}
M_{\nu}^{D} = 
\renewcommand{\arraystretch}{1.3}
\left(
\begin{matrix}
\frac{v_{H}}{\sqrt{2}} U Y_{\nu}^{\rm diag}  &   -f_{\nu}\, \lambda_{\ell}   \\
0  &  m'_{\nu_{\tau}}  \\
\end{matrix}
\right) \, .
\end{equation}
If $\mathcal{M}_{\nu}$ has the ISS hierarchy $\widetilde{\mu}_{R}, \mu_{S} \ll m_{\nu}^{D} < \widetilde{M}_{R}$, then there are three light Majorana neutrinos with a mass matrix of the form~\footnote{Here, we have taken $\widetilde{\mu}_{R} = {\bf 0}$ for simplicity, but its inclusion does not change the effectiveness of the ISS mechanism if it obeys the hierarchy $\widetilde{\mu}_{R}, \mu_{S} \ll m_{\nu}^{D} < \widetilde{M}_{R}$~\cite{Abada:2014vea,Dev:2012sg}. The same is true {for} lepton number violating couplings of the form $ \, \overline{\ell}'_{L}\widetilde{H} S_{R}$ which are in principle allowed by gauge invariance~\cite{Abada:2014vea,Dev:2012sg,CarcamoHernandez:2017owh}.}
\begin{equation}
M_{\rm light} \approx M_{\nu}^{D} \widetilde{M}_{R}^{-1} \mu_{S} \, (\widetilde{M}_{R}^{T})^{-1} (M_{\nu}^{D})^{T} \,,
\label{ISS_light_M}
\end{equation}
and six heavy Majorana states which can be grouped into three pairs with mass splittings proportional to $\mu_{S}$, such that they behave as three heavy pseudo-Dirac neutrinos with masses $\mathcal{O}(\widetilde{M}_{R})$~\cite{Dias:2012xp,Abada:2014vea}.
In the ISS limit, sub-eV masses can be achieved for the light Majorana neutrinos even if the Dirac mass is $\mathcal{O}(v_{\rm EW})$ and $\widetilde{M}_{R}$ is $\mathcal{O}(v_{1})$, as long as $\mu_{S}$ is very small. 
{The fields $\nu_{R}'^{i}$, $\nu_{R}'^{3}$, and $S_{R}$ carry $U(1)_{L}$ number 1, 1, and -1, respectively, so the two terms in the first line of $\mathcal{L}_{\nu}$ in Eq.~\eqref{YukL} are $L$ conserving, whereas the terms with $\mu_{R}$ and $\mu_{S}$ in the second line violate $L$ by 2 units. Thus, it is natural in the t'Hooft sense~\cite{tHooft:1979rat} for $\mu_{R}$ and $\mu_{S}$ to be small parameters because $U(1)_{L}$ symmetry is restored in the limit that  $\mu_{R}, \mu_{S} \rightarrow 0$. In this limit, the six heavy Majorana states become three heavy exactly Dirac neutrinos and the three active Majorana neutrinos become exactly massless because the $U(1)_{L}$ symmetry forbids Majorana mass terms. 

%%%%%%%%%%%%%%%%%
\section{Gauge Interactions and Phenomenology} 
%%%%%%%%%%%%%%%%%

\subsection{Fermion Interactions with Gauge Bosons}

Let us denote fermion representations with multiple flavor copies under the unbroken SM gauge group $G_{SM}$ as ${\bf Q}'_L = (q'^i_L, q'^3_L, Q'_L)^T$, and ${\bf L}'_L  = (\ell'^i_L, \ell'^3_L, L'_L)^T$ for left-handed fields, and ${\bf U}'_R  = (u'^i_R, u'^3_R)^T$, ${\bf D}'_R = (d'^i_R, d'^3_R)^T$, and ${\bf E}'_R  = (e'^i_R, e'^3_R)^T$ for right-handed fields, where $i = 1,2$. Expanding the kinetic terms of the fermions leads to the following $Vff$ couplings in the interaction basis:
{\footnotesize
\begin{align}
&\mathcal{L}_{U_1} \supset \frac{g_4}{\sqrt{2}} \left(\bar{Q}'_L \gamma^\mu {L}'_L + \bar q'^3_L \gamma^\mu \ell'^3_L \right) \, U_{1 \mu}~\nonumber \\
&+\frac{g_4}{\sqrt{2}} \left( \bar{Q}'_R \gamma^\mu L'_R + \bar d'^3_R \gamma^\mu e'^3_R +\bar u'^3_R \gamma^\mu \nu'^3_R \right) \, U_{1 \mu} + \textrm{h.c.}~, \label{eq:u1lag}\\
&\mathcal{L}_{g'} \supset  \frac{g_4 g_s}{g_3} \left( \bar{{\bf Q}}'_L \, {\it C}^L_{g'} \gamma^\mu T^a {\bf Q}'_L + \bar{Q}'_R \gamma^\mu T^a Q'_R \right) g'^a_\mu ~\nonumber \\
&+  \frac{g_4 g_s}{g_3} \left( \bar{{\bf U}}'_R \, {\it C}^R_{g'} \gamma^\mu T^a {\bf U}'_R +  \bar{{\bf D}}'_R \, {\it C}^R_{g'} \gamma^\mu T^a {\bf D}'_R \right) g'^a_\mu ~,\label{eq:gplag}\\
&\mathcal{L}_{Z'} \supset \frac{\sqrt{3} \,g_4 g_Y}{6 \sqrt{2} \,g_1} \left( \bar{{\bf Q}}'_L \, {\it C}_{Z'}^L \gamma^\mu {\bf Q}'_L - 3\,  \bar{{\bf L}}'_L \, {\it C}_{Z'}^L \gamma^\mu {\bf L}'_L \right) Z'_\mu ~\nonumber\\
&+\frac{\sqrt{3} \,g_4 g_Y}{6 \sqrt{2} \,g_1} \left( \bar{ Q}'_R \gamma^\mu  Q'_R - 3\,  \bar{ L}'_R \gamma^\mu L'_R \right) Z'_\mu ~\nonumber\\
&+\frac{\sqrt{3} \,g_4 g_Y}{6 \sqrt{2} \,g_1}\left( \bar{{\bf U}}'_R \, {\it C}_{Z'}^U \gamma^\mu {\bf U}'_R +  \bar{{\bf D}}'_R \, {\it C}_{Z'}^D \gamma^\mu {\bf D}'_R \right) Z'_\mu ~\nonumber\\
&-\frac{\sqrt{3} \,g_4 g_Y}{2 \sqrt{2} \,g_1}\left(\bar{{\bf E}}'_R \, {\it C}_{Z'}^E \gamma^\mu {\bf E}'_R -  \left(1 + \frac{2 g_1^2}{3 g_4^2}\right) \, \bar{\nu}'^3_R \gamma^\mu \nu'^3_R \right) Z'_\mu ~, \label{eq:zplag}
\end{align}}
\!\!where
{\footnotesize
\begin{align}
&{\it C}_{g'}^L\!\!=\!\diag\!\!\left(\!\frac{\text{-}g_3^2}{g_4^2}, \frac{\text{-}g_3^2}{g_4^2},1,1\!\right)\!,\,
{\it C}_{Z'}^L\!\!=\!\diag\!\!\left(\frac{\text{-}2 g_1^2}{3 g_4^2}, \frac{\text{-}2 g_1^2}{3 g_4^2},1,1\!\right),\nonumber\\
&{\it C}_{g'}^R\!\!=\!\diag\!\!\left(\!\frac{\text{-}g_3^2}{g_4^2}, \frac{\text{-}g_3^2}{g_4^2},1\!\right)\!,\,
{\it C}_{Z'}^U\!\!=\!\diag\!\!\left(\!\frac{\text{-}8 g_1^2}{3 g_4^2}, \frac{\text{-}8 g_1^2}{3 g_4^2},1\!-\!\frac{2 g_1^2}{g_4^2}\!\right),\nonumber\\
&{\it C}_{Z'}^D\!\!=\!\diag\!\!\left(\!\frac{4 g_1^2}{3 g_4^2},\!\frac{4 g_1^2}{3 g_4^2},\!1\!+\!\frac{2 g_1^2}{g_4^2}\!\right)\!,
{\it C}_{Z'}^E\!\!=\!\diag\!\!\left(\!\frac{\text{-}4 g_1^2}{3 g_4^2},\!\frac{\text{-}4 g_1^2}{3 g_4^2},\!1\!-\!\frac{2 g_1^2}{3 g_4^2}\!\right).
\end{align}}
\!\!Note that the right-handed representations $Q'_R$ and $L'_R$ come in a single copy of flavor. The relevant interactions in the mass basis are obtained after applying the appropriate rotation matrices.

A detailed phenomenological survey of the model is beyond the scope of the present work. Here, we comment only on a few interesting effects in low- and high-$p_T$ experiments. A good example is the LHC phenomenology in the limit $g_4 \gg g_1, g_3$, where one finds $g_3 \approx g_s$ and $g_1 \approx g_Y$. In this case, the $g'$ and $Z'$ bosons decay dominantly to a pair of third family SM fermions (or to $\chi$ if it is light enough) and the production cross section in $pp$ collisions for $g'$ and $Z'$ from the valence quarks is suppressed, relaxing otherwise strong bounds. This also makes direct searches for the $U_1^\mu$ vector LQ~\cite{Sirunyan:2017yrk,Aad:2015caa} relevant because the mass spectrum of the gauge bosons cannot be significantly split~\cite{DiLuzio:2017vat}. The present LHC limits on these states are already $\gtrsim$\,TeV, with significant prospect for improvements in the future.\footnote{For a recent review on LQ physics, see~\cite{Dorsner:2016wpm}. LQ direct search phenomenology at hadron colliders was recently reviewed in~\cite{Dorsner:2018ynv}.}

When integrated out, these vector resonances lead to four-fermion operators, which could give an observable indirect signal in low-energy flavor and electroweak precision observables. However, thanks to the approximate $U(2)$ flavor symmetry, the rotation matrices which control flavor violation are close to identity and exhibit enough suppression~\cite{Barbieri:2011ci,Barbieri:2012uh} to allow for TeV scale vector resonances.\footnote{The effects of scalar resonances in flavor physics are typically further suppressed by the light fermion masses.} Flavor effects of the color octet in this context have been discussed in Ref.~\cite{Chivukula:2013kw}. Important constraints come from the neutral meson oscillation phenomena in the down quark sector (e.g. $B_s$-$\bar B_s$ mixing), effectively requiring down-alignment~\cite{DiLuzio:2017vat}.  A TeV scale vector LQ with left-handed interactions controlled by an approximate $U(2)$ flavor symmetry has been shown to be compatible with the constraints from semi-leptonic and rare meson decays, LFU and LFV in charged lepton decays, and $Z$ and $W$-pole precision measurements (see e.g. Ref.~\cite{Barbieri:2015yvd,Buttazzo:2017ixm}). It is crucial to note that in the limit of large $m_{\chi}$, this model is the low energy limit of the $PS^3$ model presented in Ref.~\cite{Bordone:2017bld}, apart from the neutral lepton sector.
 
\subsection{$B$-Physics Anomalies and PMNS Non-unitarity}

The model proposed here can accommodate the recently reported anomalies in $B$-meson decays both in (i) deviations from $\tau / \ell$ (where $\ell= e,\mu$) universality in semi-tauonic decays as defined by $R(D^{(*)})$ observables (charged $b \to c \ell \nu$ transitions)~\cite{Lees:2013uzd,Hirose:2016wfn,Aaij:2015yra} and (ii) deviations from $\mu / e$ universality in rare decays as defined by $R(K^{(*)})$ observables (neutral $b \to s \ell \ell$ transitions)~\cite{Aaij:2014ora,Aaij:2017vbb}. Basically, the $U_1^\mu$ vector LQ induces a large tree-level contribution to $b \to c \tau \nu$ while simultaneously giving a flavor-suppressed tree-level contribution to $b \to s \mu \mu$. We note that the dimension-6 effective operator introduced in~\cite{Bordone:2017bld} to solve $b \to s \mu \mu$ is generated in our model when integrating out the vector-like fermion field $\chi$.

The rest of the discussion on $B$-anomalies follows Ref.~\cite{Bordone:2017bld}, and we do not repeat it here. Nonetheless, let us point to a novel correlation between $B$-anomalies and non-unitarity in the Pontecorvo-Maki-Nakagawa-Sakata (PMNS) matrix, both of which are controlled by the ratio of $G_{\textrm{SM}}$- and $G$-breaking scales. On the one hand, the non-standard contribution to $\Delta R_{D^{(*)}}^{\tau \ell} \equiv R(D^{(*)})/R(D^{(*)})_{\textrm{SM}} - 1$ is~\cite{Bordone:2017bld}
\begin{equation}
\Delta R_{D}^{\tau \ell} \approx 2.2 \, \Delta R_{D^*}^{\tau \ell} \approx \frac{5 v_{\rm EW} ^2}{v_1^2 + v_3^2}~.
\end{equation}
Since $\vev{\Omega_3} (\equiv v_3 /\sqrt{2} )$ controls the mass of the coloron field $g'$, it is bounded from below by direct searches at the LHC ($v_3 \gtrsim 1$~TeV). On the other hand, deviation of the PMNS matrix from unitarity is parameterized by the Hermitian matrix $\epsilon = {\bf 1} - NN^{\dagger}$, where $N$ is the non-unitarity PMNS matrix. In terms of the ISS parameters introduced in Section~\ref{lep_sec}, we can write $\epsilon$ approximately as~\cite{Dias:2012xp}
\begin{equation}
\epsilon  \approx (M_{\nu}^{D})^{*} (\widetilde{M}_{R}^{-1})^{*} (\widetilde{M}_{R}^{-1})^{T} (M_{\nu}^{D})^{T} \, .
\end{equation}
If we require $v_{3} \gtrsim 1$ TeV in order to evade the bound on direct searches for the coloron, we then require $v_{1} \lesssim 1$ TeV in order to produce the observed anomaly in $\Delta R_{D}^{\tau \ell}$. Thus, there is a contribution to $\epsilon$ which is a least as large as
\begin{equation}
\epsilon \sim \frac{v_{\rm EW}^{2}}{v_{1}^{2} |\lambda_{R}|^{2}} \, ,
\label{eq:eps_min}
\end{equation}
meaning that significant PMNS unitarity violation is associated with a quark-lepton unification scale which is low enough to explain $\Delta R_{D}^{\tau \ell}$. The two ways to avoid the unitarity bound are: a large coupling $|\lambda_{R}|$ or accepting some tuning in $M_{\nu}^{D}$. {For example, the benchmark point shown in Appendix~\ref{app:A} predicts PMNS unitarity violation which is just below the present limits with $M_{\nu}^{D}\sim 10^{-1} v_{\rm EW}$, however, a large coupling $|\lambda_{R}| \sim 3$ is required.}

%%%%%%%%%%%%%%%%%
 \section{Conclusions} 
%%%%%%%%%%%%%%%%%
%
We have constructed a model of TeV scale quark-lepton unification based on an extended ``4321" gauge group, where the third family quarks and leptons are unified into fundamental representations of $SU(4)$ while the light family fermions are charged only under $``321"$. As a result of this construction, the model contains an accidental $U(2)$ flavor symmetry which suppresses FCNC and allows for the realization of the correct CKM texture.  A key prediction of the model is a gauge vector leptoquark $U_{1}^{\mu}$, coupled dominantly to the third family, and potentially within reach of the LHC.

While third family quark-lepton unification nicely explains the closeness of the tau lepton and bottom quark masses, it fails spectacularly in the up sector, suggesting a peculiar origin for neutrino masses. In particular, the model, with the addition of gauge singlet fermions, admits a natural realization of light neutrino masses via the inverse seesaw mechanism. In this article, we present a numerical benchmark point where experimentally acceptable masses and mixings are obtained for the light neutrinos. 
%without fine-tuning.

This model is a very interesting and phenomenologically rich construction, predicting a plethora of observable effects ranging from low energy neutrino and flavor physics up to high-$p_{T}$ collider searches. We may already be seeing its first signatures in the still inconclusive $B$-anomalies.

%%%%%%%%%%%%%%%%%

\section*{Acknowledgments}
We would like to thank Luca Di Luzio for pointing us to Ref.~\cite{Perez:2013osa}, and Joachim Kopp for reading the manuscript carefully. We also thank Javier Fuentes Martin, Marco Nardecchia, and Toby Opferkuch for useful discussions. Finally, we thank Deutsche Bahn for their hospitality, during which the initial idea for this manuscript was conceived. The results of this article were already publicly presented at the Zurich Phenomenology Workshop (ZPW2018) by one of us.

\appendix
\section{Neutrino masses, mixings, and PMNS Non-Unitarity}
\label{app:A}
\subsection{Light Neutrino Masses and Mixings}
In Section~\ref{lep_sec}, we wrote the neutrino mass Lagrangian as
\begin{equation}
\mathcal{L}_{\nu} = -\frac{1}{2} \overline{n}_{L} \mathcal{M}_{\nu} n_{L}^{c} + \textrm{h.c.}\, ,
\end{equation}
where the neutrino mass matrix is a $9 \times 9$ complex symmetric matrix which has the ISS texture
\begin{equation}
\mathcal{M}_{\nu} = 
\renewcommand{\arraystretch}{1.3}
\left(
\begin{matrix}
0 &   M_{\nu}^{D} & 0  \\
(M_{\nu}^{D})^{T}  &  \widetilde{\mu}_{R} & \widetilde{M}_{R}^{T}  \\
0 & \widetilde{M}_{R} & \mu_{S}\\
\end{matrix}
\right) \, .
\label{eq:ISS_matrix_app}
\end{equation}
In terms of the original Lagrangian parameters, the mass matrix has the form
{\footnotesize
\begin{equation}
\mathcal{M}_{\nu} = 
\renewcommand{\arraystretch}{1.3}
\left(
\begin{matrix}
0  &  0 &\frac{v_{H}}{\sqrt{2}} U Y_{\nu}^{\rm diag}  &  -f_{\nu}\, \lambda_{\ell}  & 0 \\
  0 &  0 & 0 & m_{\nu_{\tau}}'  & 0 \\
 \frac{v_{H}}{\sqrt{2}} Y_{\nu}^{\rm diag}U^{T}   &  0  & \mu_{R} &  0& M_R^{T}  \\
    -f_{\nu}\, \lambda_{\ell}^{T}  &   m_{\nu_{\tau}}' &  0 & 0  & \frac{v_{1}}{\sqrt{2}}\lambda_{R}^{T} \\
  0 & 0 & M_R &\frac{v_{1}}{\sqrt{2}}\lambda_{R}   & \mu_{S} \\  
\end{matrix}
\right) \,.
\label{eq:9by9_neu}
\end{equation}
}
The entry $\frac{v_{H}}{\sqrt{2}}UY_{\nu}^{\rm diag}$ is the product of a $2\times2$ orthogonal matrix with a $2\times2$ real diagonal matrix which we parameterize as
\begin{equation}
\frac{v_{H}}{\sqrt{2}}UY_{\nu}^{\rm diag} = 
\renewcommand{\arraystretch}{1.3}
\left(
\begin{matrix}
\cos\theta  &    \sin\theta   \\
-\sin\theta  &  \cos\theta  \\
\end{matrix}
\right) 
\left(
\begin{matrix}
m'_{\nu_{e}}  &    0  \\
0  & m'_{\nu_{\mu}} \\
\end{matrix}
\right)\, .
\end{equation}

To reduce the number of free parameters, we seek a solution which yields acceptable light neutrino masses and mixings with a simplifying ansatz where $\mu_{R} = {\bf0}$, $\mu_{S} = \diag(\mu_1\,, \mu_2 \,, \mu_{3})$, $\lambda_{\ell}^{T} = (\lambda_{\ell}^{(1)}, \, \lambda_{\ell}^{(2)})$ and
\begin{equation}
M_{R} = 
\renewcommand{\arraystretch}{1.3}
\left(
\begin{matrix}
m_{R}  &   0   \\
0  &  m_{R}  \\
0  &  0  \\
\end{matrix}
\right) \, , \hspace{5mm}
\frac{v_{1}}{\sqrt{2}}\lambda_{R} =  
\left(
\begin{matrix}
0  \\
0  \\
m_{QL}  \\
\end{matrix}
\right)\, ,
\end{equation}
such that $ \widetilde{M}_{R} = (M_{R} \,\,\, \frac{v_{1}}{\sqrt{2}}\lambda_{R})$ is a diagonal $3\times 3$ block of $\mathcal{M}_{\nu}$. Here, $m_{QL}$ is of order the $SU(4)$  breaking  (quark-lepton unification) scale. We also take all parameters to be real. Following the prescription in Refs~\cite{Schechter:1981cv,Hettmansperger:2011bt,Dias:2012xp}, we first block diagonalize $\mathcal{M}_{\nu}$ via a rotation $\mathcal{W}$ such that
\begin{equation}
\mathcal{W}^{T}\mathcal{M}_{\nu} \mathcal{W} 
= \left(
\begin{matrix}
(M_{\rm light})_{3\times3} &    0_{3\times6}  \\
0_{6\times3}  & (M_{\rm heavy})_{6\times6} \\
\end{matrix}
\right)\, .
\end{equation}
The rotation matrix $\mathcal{W}$ is approximately given as
\begin{equation}
\mathcal{W}
\approx \left(
\begin{matrix}
\left[{\bf 1}-\frac{1}{2} \Theta\Theta^{\dagger}\right]_{3\times3} &    \Theta_{3\times6}  \\
\Theta^{\dagger}_{6\times3}  & \left[{\bf 1}-\frac{1}{2} \Theta\Theta^{\dagger}\right]_{6\times6} \\
\end{matrix}
\right)\, ,
\end{equation}
assuming $\Theta$ is given as a power series in $M_{\rm heavy}^{-1}$ as $\Theta = \sum_{i} \Theta_{i}$ with $\Theta_{i} \sim (M_{\rm heavy}^{-1})^{i}$. In terms of the ISS parameters in Eq.~\ref{eq:ISS_matrix_app}, the $3 \times 3$ mass matrix for the light Majorana neutrinos $M_{\rm light}$ has the usual ISS form
\begin{equation}
M_{\rm light} \approx M_{\nu}^{D} \widetilde{M}_{R}^{-1} \mu_{S} \, (\widetilde{M}_{R}^{T})^{-1} (M_{\nu}^{D})^{T} \,.
\label{eq:mL}
\end{equation}
If we now diagonalize $M_{\rm light}$ and $M_{\rm heavy}$ via rotations of the form $U_{l}$ and $U_{h}$, the complete matrix $\mathcal{U}$ which diagonalizes $M_{\nu}$ is 
\begin{equation}
\mathcal{U}
\approx \left(
\begin{matrix}
\left[{\bf 1}-\frac{1}{2} \Theta\Theta^{\dagger}\right] U_{l} &    \Theta \, U_h  \\
\Theta^{\dagger} \, U_l & \left[{\bf 1}-\frac{1}{2} \Theta\Theta^{\dagger}\right] U_h \\
\end{matrix}
\right)\, ,
\end{equation}
and the light neutrino flavor eigenstates are given by
\begin{equation}
\nu_{\alpha} \approx \left[ U_{l} - \frac{1}{2} \Theta_{1}\Theta_{1}^{\dagger}U_{l} \right]_{\alpha i} \, \hat{n}_{l}^{i} + \left[\Theta_{1} U_h\right]_{\alpha \, b} \, \hat{n}_{h}^{b} \, ,
\end{equation}
where $\alpha = e,\mu,\tau$, $\hat{n}_{l}^{i}$ are the light mass eigenstates with $i=1,..,3$, and $\hat{n}_{h}^{b}$ are the heavy mass eigenstates with $b=1,..,6$.
The PMNS neutrino mixing matrix is now a non-unitary matrix given by
\begin{equation}
N = \left[ {\bf 1} - \frac{1}{2} \Theta_{1}\Theta_{1}^{\dagger}\right]U_{l} \, ,
\label{eq:non-uni-PMNS}
\end{equation}
where $\epsilon \equiv {\bf 1} - NN^{\dagger}  \approx \Theta_{1}\Theta_{1}^{\dagger}$ parameterizes the deviation of $NN^{\dagger}$ from unitarity. The $3\times 3$ Hermitian matrix $\epsilon$ can be written approximately in terms of the original ISS parameters as 

\begin{equation}
\epsilon  \approx (M_{\nu}^{D})^{*} (\widetilde{M}_{R}^{-1})^{*} (\widetilde{M}_{R}^{-1})^{T} (M_{\nu}^{D})^{T}
\end{equation}

\begin{table}[h!]
\renewcommand{\arraystretch}{1.1}
\begin{center}
\begin{tabular}{|c|c|}
\hline 
ISS Parameter  & Value     \\ 
\hline 
$m_{\nu_{e}}'$  & 1.67 GeV     \\ 
$m_{\nu_{\mu}}'$  & 38.3 GeV     \\ 
$m_{\nu_{\tau}}'$  & 10.0 GeV     \\
\hline
$\sin\theta$  & 0.510    \\
$f_{\nu} \lambda_{\ell}^{(1)}$  & 0.883 GeV   \\
$f_{\nu} \lambda_{\ell}^{(2)}$  & 6.80 GeV   \\
\hline
$m_{QL}$  & 2.00 TeV     \\ 
$m_{R}$  & 10.0 TeV     \\ 
\hline
$\mu_{1}$  & 0.720 keV     \\
$\mu_{2}$  & 0.871 keV     \\
$\mu_{3}$  & 1.28 keV     \\
  \hline
\end{tabular}
\end{center}
\caption{ISS parameters for a simplified benchmark point.} 
\label{tab:ISS_fit}
\end{table}

Assuming the simplifying ansatz for the ISS parameters outlined in the previous section, we diagonalize $M_{\rm light}$ numerically using the benchmark parameter set in Table~\ref{tab:ISS_fit}. We obtain a normal hierarchy of light neutrino masses of $m_{1} \approx 1.86\times10^{-2}$ meV, $m_{2} \approx 8.58$ meV,  $m_{3} \approx 51.3$ meV with mass-squared splittings of
\begin{align}
&\Delta m_{32}^{2} = 2.56 \times 10^{-3} \,\, {\rm eV^{2}} \, ,~\nonumber \\
&\Delta m_{21}^{2} = 7.36 \times 10^{-5} \,\, {\rm eV^{2}} \, .
\end{align}
To construct the PMNS matrix, we numerically find the matrix $U_l$ which diagonalizes $M_{\rm light}$ as $M_{\rm light}^{\rm diag} = U_{l}^{T} M_{\rm light} U_{l}$ and use Eq.~\ref{eq:non-uni-PMNS}. We obtain the following results for the mixing angles
\begin{align}
&\sin^{2}\theta_{12} = 0.296 \, ,~\nonumber \\
&\sin^{2}\theta_{23} = 0.425 \, ,~\nonumber \\
&\sin^{2}\theta_{13} =  0.0214 \, .
\end{align}

{These mass-squared splittings and mixing angles agree very well with the best-fit values derived from a global fit of the current neutrino oscillation data in Refs~\cite{Patrignani:2016xqp,Capozzi:2016rtj}. We have also performed an exact numerical diagonalization of Eq.~\ref{eq:9by9_neu} and found very good agreement with the approximate masses and mixings given by diagonalizing Eq.~\ref{eq:mL}.
More general benchmark points are of course possible if the simplifying assumptions about $\mathcal{M}_{\nu}$ made here are relaxed, e.g. leaving $\widetilde{M}_{R}$ as a general complex $3\times 3$ matrix. One can even consider non-equal numbers of $\nu_{R}'$ and $S_{R}$ in order to have additional sterile neutrino states at the scale $\mu_S$~\cite{Abada:2014vea,Aguilar:2001ty,Armbruster:2002mp,AguilarArevalo:2008rc,AguilarArevalo:2010wv,Mention:2011rk,Kopp:2011qd}.}

%which may be of interest in the context of anomalies in short baseline neutrino oscillation experiments~\cite{Aguilar:2001ty,Armbruster:2002mp,AguilarArevalo:2008rc,AguilarArevalo:2010wv,Mention:2011rk,Kopp:2011qd}. 
%However, a detailed scan of the ISS parameters is beyond the scope of this work.

\subsection{PMNS Non-Unitarity}
We quantify the deviation of the PMNS matrix from unitarity by $|\epsilon| = |{\bf 1} - NN^{\dagger}|$. For the benchmark point, the matrix $|\epsilon|$ is 
\begin{equation}
|\epsilon| = 
\renewcommand{\arraystretch}{1.3}
\left(
\begin{matrix}
4.04\times 10^{-6} & 7.94\times 10^{-6} & 2.21\times 10^{-6}  \\
7.94\times 10^{-6} & 2.24\times 10^{-5} & 1.70 \times 10^{-5} \\
2.21\times 10^{-6} & 1.70 \times 10^{-5} & 2.50\times 10^{-5} \\
\end{matrix}
\right) \, .
\label{eq:eta_num}
\end{equation}
The largest source of non-unitarity is coming from $|\epsilon_{33}|$ for which (assuming the benchmark parameters) a simple analytic approximation can be obtained
\begin{equation}
|\epsilon_{33}| \approx \left(\frac{m_{\nu_{\tau}}'}{m_{QL}}\right)^{2} \,,
\end{equation}
which is in agreement with the estimate in Eq.~\ref{eq:eps_min}.
 The current bounds on PMNS non-unitarity are \cite{Antusch:2014woa,Antusch:2015mia}
\begin{equation}
|\epsilon| < 
\renewcommand{\arraystretch}{1.3}
\left(
\begin{matrix}
2.1\times 10^{-3} & 1.0\times10^{-5}& 2.1\times 10^{-3}  \\
1.0\times10^{-5} & 4.0\times 10^{-4} & 8.0 \times 10^{-4} \\
2.1\times 10^{-3} & 8.0 \times 10^{-4} & 5.3\times 10^{-3} \\
\end{matrix}
\right) \, .
\label{eq:eta_num}
\end{equation}

 \bibliographystyle{apsrev4-1.bst}
  \bibliography{bibliography}

\end{document}